\documentclass{article}
\usepackage{spconf,amsmath,graphicx}
\usepackage{booktabs}
\usepackage{caption}
\usepackage{subcaption}
\usepackage{dsfont}
\usepackage{bm}
\usepackage{setspace}
\usepackage{hyperref}

\title{Bayesian adaptive learning to latent variables via Variational Bayes and Maximum a Posteriori}
\name{Hu Hu$^{1}$,
      Sabato Marco Siniscalchi$^{1,2}$,
      Chin-Hui Lee$^{1}$
      }
\address{$^1$School of Electrical and Computer Engineering, Georgia Institute of Technology, GA, USA \\
$^2$Computer Engineering School, University of Enna Kore, Italy\\
}
\begin{document}
%
\maketitle
\begin{abstract}
In this work, we aim to establish a Bayesian adaptive learning framework by focusing on estimating latent variables in deep neural network (DNN) models. Latent variables indeed encode both transferable distributional information and structural relationships.  Thus the distributions of the source latent variables (prior) can be combined with the knowledge learned from the target data (likelihood) to yield the distributions of the target latent variables (posterior) with the goal of addressing acoustic mismatches between training and testing conditions. 
The prior knowledge transfer is accomplished through Variational Bayes (VB). In addition,  we also investigate Maximum a Posteriori (MAP) based Bayesian adaptation. 
Experimental results on device adaptation in acoustic scene classification show that our proposed approaches can obtain good improvements on target devices, and consistently outperforms other cut-edging algorithms.
\end{abstract}
\section{Introduction}
\label{sec:intro}


Bayesian learning has achieved a great deal of success for model adaptation \cite{ripley1996pattern, lee2000adaptive, huang2017bayesian}. It provides a mathematical framework to model uncertainties and take into account prior knowledge of the problem. 
Specifically, by leveraging upon some target data and a prior belief, a posterior belief can be optimally established. A common adaptation mechanism is the maximum a posteriori (MAP) estimation of parameters in the posterior distributions, which amounts to maximizing the posterior probability and gives us a point estimation of the model parameters. It has been proven effective in handling acoustic mismatches in hidden Markov models (HMMs) \cite{lee2000adaptive, gauvain1994maximum, siohan2001joint} and DNNs \cite{huang2015maximum, huang2017bayesian} by assuming a distribution on the model parameters.  The entire posterior distributions can also be approximated by other Bayesian approaches, such as Markov Chain Monte Carlo (MCMC), assumed density filtering, and variational Bayes (VB).  In particular, VB transforms an estimation problem into an optimization one that can be solved numerically by leveraging a variational distribution. It has been extensively adopted in conventional acoustic models \cite{watanabe2004variational} and deep architectures \cite{kingma2013auto, hu2022variational}.
The Bayesian learning framework is known to facilitate constructing an adaptive system for specific target conditions in a particular environment. Thus the mismatch between training and testing can be reduced with an overall improvement of system performance.

In this work, we aim to establish a Bayesian adaptive learning framework focusing on estimating DNN latent variables,  with the goal of bridging the performance gap between source and target domains. Those latent variables refer to the unobservable representations of the data, and deep latent variables usually correspond to intermediate embedding from a hidden specific layer. Specifically, the hidden embedding of a well-trained DNN model usually has its own distribution form. Each hidden embedding can be regarded as a sampled instance of the deep latent variables. Thus in particular, the distributions of the source latent variables (prior) are combined with the knowledge learned from the target data (likelihood) to yield the distributions of the target latent variables (posterior). Prior knowledge is thus encoded and transferred from the source to the target domains. In \cite{hu2022variational}, Bayesian adaptation via variational Bayes is proposed. In this work, we further proposed and investigated MAP adaptation. 
Gaussian and Dirichlet assumptions are cast to approximate the posterior distribution. The proposed methods are evaluated on device adaptation tasks in acoustic scene classification. Experimental results show that our proposed approaches can obtain good improvements on target devices.

\section{Bayesian Adaptation}
\label{sec:bayes}

\subsection{Bayesian Inference of Latent Variables}
\label{sec:bayes1}
Suppose we are given some observations $\mathcal{D}$, and let $\mathcal{D}_S = \{x_S^{(i)}, y_S^{(i)}\}^{N_S}_{i=1}$ and $\mathcal{D}_T = \{x_T^{(i)}, y_T^{(i)}\}^{N_T}_{i=1}$ be the source and target domain data, respectively.
Therefore,  there exists a corresponding data sample $x_S^{(j)}$ from the source data for each target data sample $x_T^{(i)}$, where $x_T^{(i)}$ and $x_S^{(j)}$ have strong relationships, e.g., the same audio content but recorded by different devices. The data samples are thereby parallel. 
Suppose to have a DNN-based model with parameters $\lambda$, i.e., weights, that need to be estimated. Starting from the classical Bayesian approach, a prior distribution $p(\lambda)$ is defined over $\lambda$, and the posterior distribution after seeing the observations $\mathcal{D}$ can be obtained by the Bayes rule as $p(\lambda|\mathcal{D}) = \frac{p(\mathcal{D}|\lambda) p(\lambda)}{p(\mathcal{D})}$.

In addition to the network weights, we also introduce the latent variables $Z$ to model the intermediate hidden embedding of the DNN. $Z$ refers to the unobserved intermediate representations, which usually encode transferable distributional information and structural relationships among data.  More specifically, $Z$ can correspond to a vector of either the intermediate or final output of a DNN. When $Z$ represents the logits, i.e., the output before the softmax, or the soft output, i.e., the output after the softmax, the mapping network from $Z$ to $Y$ has no-trainable parameters.
Next, we decouple the network weights into two independent subsets, $\theta$ and $\omega$, to represent weights before and after $Z$, respectively. Thus we have $p(\lambda) = p(Z, \theta, \omega) = p(Z|\theta) p(\theta) p(\omega)$.
This relationship holds for both prior $p(\lambda)$ and posterior $p(\lambda|\mathcal{D})$. Here we focus on transferring knowledge in a distribution sense via the latent variables $Z$ leveraging parallel data. Therefore, we assume there exists $Z$ retaining the same distribution across the source and target domains. $\lambda_S = \{Z_S, \theta_S, \omega_S\}$, and $\lambda_T =  \{Z_T, \theta_T, \omega_T\}$ are used for source and target domains, respectively. For the target model, we have the prior knowledge learned from the source as $p(Z_T|\theta_T) = p(Z_S|\theta_S, \mathcal{D}_S)$. The posterior $p(\lambda_T|\mathcal{D}_T)$ is usually intractable, and an approximation is required. There are many ways to perform the approximation, like maximum a posteriori (MAP), variation Bayes (VB), and Markov Chain Monte Carlo (MCMC). We proposed a VB-based framework in \cite{hu2022variational}. In this work, we also investigate a MAP-based solution. 

\subsection{Bayesian Adaptation via Variational Bayes}
We here describe Bayesian adaptation via variational Bayes (BA-VB), which is proposed in \cite{hu2022variational} and named as VBKT, a variational distribution $q(\lambda_T|\mathcal{D}_T)$ is set to approximate the posterior. Specifically, for the target domain model, the optimal $q^*(\lambda_T|\mathcal{D}_T)$ is obtained by minimizing KL divergence between the variational distribution and the real one, over a family of allowed approximate distributions.
As we focus on latent variables, $Z$, a non-informative prior is assumed over $\theta_T$ and $\omega_T$. For the variational distribution, a Gaussian mean-field approximation is used to specify the distribution forms for both the prior and posterior over $Z$. By introducing prior distribution, as induced in \cite{hu2022variational}, the final objective function is: 
\begin{align}
    \mathcal{L}(\lambda_T; \mathcal{D}_T) =\ & \sum^{N_T}_i \mathds{E}_{z^{(i)}_T \sim \mathcal{N}(\mu_T^{(i)}, \sigma^2)} \log p(y_T^{(i)} | x_T^{(i)}, z^{(i)}_T, \theta_T, \omega_T)\nonumber \\
    & - \frac{1}{2\sigma^2} \sum_i^{N_T} \| \mu_T^{(i)} - \mu_S^{(i)}\|_2^2,
    \label{eq:vb-obj}
\end{align}
where the first term is the likelihood, and the second term is deduced from the KL divergence between prior and posterior of the latent variables. Each instance of $z^{(i)}_T$ is sampled from the posterior distribution as $z_T^{(i)} \sim \mathcal{N}(\mu_T^{(i)}, \sigma^2)$, and the expectation can be reduced via the reparameterization trick \cite{kingma2013auto}.

\subsection{Bayesian Adaptation via Maximum a Posteriori}

Instead of considering the full posterior distribution, we here introduced MAP to approximate the prosterior via a point estimation. 
MAP generates a point estimate of our parameters but takes into account prior knowledge.
By using relationship between $\lambda, Z, \theta, \omega$ described in \ref{sec:bayes1}, and$p(Z_T|\theta_T) = p(Z_S|\theta_S, \mathcal{D}_S)$, $\lambda_T^*$ is obtained as follows:
\begin{align}
    \lambda_T^* & = \mathop{argmax}\limits_{\lambda_T} p(\lambda_T|\mathcal{D}_T) \nonumber\\
    & = \mathop{argmax}\limits_{\theta_T, \omega_T} \log p(\mathcal{D}_T |\theta_T, \omega_T) + \log p(Z_S|\theta_S, \mathcal{D}_S), 
    \label{eq:map}
\end{align}
where the first term is the likelihood and the second term represents the prior knowledge transferred from the source domain. Similarly to BA-VB,  $Z$ is assumed to be Gaussian. Therefore, $z$ in $Z$ follows an isotropic Gaussian distribution with component weights $\alpha$, which is assumed to be the same for simplicity. We denote the Gaussian mean and variance for the source and target domains as $\mu_S; \sigma_S^2$ and $\mu_T; \sigma_T^2$, respectively. The final objective function is:
\begin{align}
    \mathcal{L}(\lambda_T; \mathcal{D}_T) =\ & \sum^{N_T}_i \log p(y_T^{(i)} | x_T^{(i)}, \theta_T, \omega_T)\nonumber \\
    & - \alpha \sum_i^N \frac{1}{2(\sigma_S^{(i)})^2} \| z_T^{(i)} - \mu_S^{(i)}\|_2^2. 
    \label{eq:map-obj-gaussian}
\end{align}
Note that for BA-MAP, different from BA-VB, $z_T^{(i)}$ here is deterministic, which is produced by input $x_T^{(i)}$ through $\theta_T$. For each unit in the second summation term, both the mean and variance can be different. If we use $z_S^{(i)}$ to represent the Gaussian mean, $z_S^{(i)} = \mu_S^{(i)}$, the second term in Eq \ref{eq:map-obj-gaussian} can be re-written to a form of mean-square-error for hidden features, with a scaling parameter sigma to control the uncertainty.

Then we discuss a case in which $Z$ represents the soft outputs (model outputs after softmax), where $\omega_S$ and $\omega_T$ are thus constants. $z^{(i)}$ here is a $K$-dims probability vector with values between [0, 1]. We can use a Dirichlet distribution, $f(z^{(i, 1)}, ..., z^{(i, K)}; \alpha^{(i, 1)}, ..., \alpha^{(i, K)}) = \frac{1}{\mathcal{B}(\mathbf{\alpha})} \prod^K_j {z^{(i, j)}}^{\alpha^{(i, j)}-1}$, rather than Gaussian distribution to model the latent variable $Z$. Thus, for a given source model, we can use the mode of Dirichlet to estimate the posterior of parameters, where $z^{(i, j)} = \frac{\alpha^{(i, j)} - 1}{\sum_j^K \alpha^{(i, j)} - K}$. If we specify a case that $\sum_j^K \alpha^{(i, j)}$ is equal to 1, we thus have $\alpha^{(i, j)} = p^{(i, j)} + 1$, where $p^{(i, j)}$ denotes each component of the probability vector $z^{(i)}$. Therefore, by reducing Eq \ref{eq:map} with Dirichlet assumption, we can finally have the objective as
\begin{align}
    \mathcal{L}(\lambda_T; \mathcal{D}_T) =\ & \sum^{N_T}_i \log p(y_T^{(i)} | x_T^{(i)}, \theta_T, \omega_T)\nonumber \\
    & - \alpha \sum_i^N \sum_j^K p_S^{(i, j)} \log p_T^{(i, j)}. 
    \label{eq:map-obj-dirichlet}
\end{align}
We can note that, in this way, Eq \ref{eq:map-obj-dirichlet} has the equivalent form to the basic teacher-student (TS) learning \cite{hinton2015distilling, li2014learning}.

\section{Experiments}
\label{sec:exp}

\subsection{Experimental Setup \& Results} 
We evaluate our proposed approaches on the acoustic scene classification (ASC) task of DCASE 2020 challenge task1a \cite{heittola2020acoustic}. The training set contains $\sim$10K scene audio clips recorded by the source device (device A), and 750 clips for each of the 8 target devices (Device B, C, s1-s6). Each target audio is paired with a source audio, and the only difference between the two audios is the recording device. The goal is to solve the device mismatch issue for one specific target device at a time, i.e., device adaptation, which is a common scenario in real applications. We adopt the same setup in \cite{hu2022variational, yen2020lottery}. An inception model is used to evaluate our proposed methods. And the latent variables are based on the hidden outputs before the last layer. For BA-VB, the same setting in \cite{hu2022variational} is utilized. For BA-MAP, we evaluated the way of using Gaussian assumption described in Section \ref{sec:bayes}. 
We carried experimental comparison against recent state-of-the-art knowledge transfer techniques. 

The experimental evaluation results are shown in \ref{tab:res-all}. The source model is trained on data recorded by Device A. There are 8 target devices, i.e., Device B, C, s1-s6. The accuracy reported in each cell of Table~\ref{tab:res-all} is obtained by averaging among 32 experimental results, from 8 target devices and 4 trials for each. As recommended in \cite{hu2022variational}, all methods are combined with the basic TS learning to obtain the best performance. Specifically, the original cross entropy (CE) loss is replaced by the addition of 0.9 $\times$ KL loss with soft labels and 0.1 $\times$ CE loss with hard labels.  The 1st row in \ref{tab:res-all} gives the results of the target model trained by target data from scratch without adaptation. The 2nd row is the result of fine-tuning the source model by target data. By comparing them we can argue the importance of knowledge transfer when building a target model. The 3rd to 7th rows show the results of compared knowledge transfer methods. Most of them show advantages over the one-hot baseline. The bottom two rows show the results of our proposed Bayesian adaptation methods, including BA-VB and BA-MAP. They not only outperform one-hot fine-tuning by a large, margin but also attain superior performance to other algorithms. From the last two rows, we can observe that BA-VB outperforms BA-MAP. As stated in \ref{sec:bayes}, BA-VB takes into account the distributions when performing the distribution estimation on latent variables, so it shows better robustness in this task.

\begin{table}[t]
\centering
\caption{Average classification accuracy (in \%)  on DCASE2020 ASC test set for different models. Each method is tested in combination with the basic TS learning method. The average is carried out over 32 trials: 8 target devices $\times$ 4 repeated trials.}
\label{tab:res-all}
\begin{tabular}{l||c}
\toprule
\toprule
\ Method      & Avg ACC (\%) \\
\midrule
\midrule
\ No transfer\ \ \ \ \ \  & 55.36        \\
\midrule
\ One-hot     & 64.38        \\
\ TSL \cite{hinton2015distilling}          & 65.71        \\
\ NLE  \cite{meng2020vector}        & 63.80        \\
\ AB  \cite{heo2019knowledge}           & 66.73        \\
\ SP   \cite{tung2019similarity}          & 66.40        \\
\ RKD  \cite{park2019relational}        & 65.16        \\
\midrule
\ BA-VB  \cite{hu2022variational}     & 69.83      \\
\ BA-MAP      & 67.88        \\
\bottomrule
\bottomrule
\end{tabular}
\end{table}



\bibliographystyle{IEEEbib}
\bibliography{strings,refs}

\end{document}